\documentclass[
    ,final            
  ]
  {aipproc}

\layoutstyle{6x9}

\begin{document}

\title{Meson wave function from \\ holographic approaches}


\author{Alfredo Vega}{
  address={Departamento de F\'\i sica y Centro de Estudios
Subat\'omicos,\\  Universidad T\'ecnica Federico Santa Mar\'\i a,\\
Casilla 110-V, Valpara\'\i so, Chile}
}

\author{Ivan Schmidt}{
  address={Departamento de F\'\i sica y Centro de Estudios
Subat\'omicos,\\  Universidad T\'ecnica Federico Santa Mar\'\i a,\\
Casilla 110-V, Valpara\'\i so, Chile}
}

\author{Tanja Branz}{
  address={Institut f\"ur Theoretische Physik,
Universit\"at T\"ubingen,\\
Kepler Center for Astro and Particle Physics, \\
Auf der Morgenstelle 14, D--72076 T\"ubingen, Germany\\}
}

\author{Thomas Gutsche}{
  address={Institut f\"ur Theoretische Physik,
Universit\"at T\"ubingen,\\
Kepler Center for Astro and Particle Physics, \\
Auf der Morgenstelle 14, D--72076 T\"ubingen, Germany\\}
}

\author{Valery~E.~Lyubovitskij}{
  address={Institut f\"ur Theoretische Physik,
Universit\"at T\"ubingen,\\
Kepler Center for Astro and Particle Physics, \\
Auf der Morgenstelle 14, D--72076 T\"ubingen, Germany\\}
}


\begin{abstract}
We discuss the light-front wave function for the valence quark
state of mesons using the AdS/CFT correspondence. 
We consider two kinds of wave functions
obtained in different holographic Soft-Wall approaches.
\end{abstract}

\maketitle

\section{Introduction}

The hadronic wave function in terms of quark and gluon degrees of freedom plays an
important role in predictions for QCD process, but in a direct extraction several problems
are encountered~\cite{BJSJK}.
For this reason there are several non-perturbative approaches to obtain properties of
distribution amplitudes and/or hadronic wave functions from QCD.
Recently new techniques based on the Anti-de Sitter / Conformal Field Theory (AdS/CFT)
correspondence have been developed, which allow to obtain mesonic 
Light-Front Wave Functions (LFWF)~\cite{BdT3, BdT4, VegaSchmidt3}.
These holographic LFWFs are  another application of Gauge/ Gravity ideas to QCD.
In fact, bottom-up models have already been quite successful in the description
of several QCD phenomena such as hadronic scattering processes~\cite{Scattering},
hadron spectra~\cite{Spectrum}, hadronic couplings and chiral symmetry breaking~\cite{Chiral}
among others.

This work is a summary of the results presented in~\cite{VegaSchmidt3} and is structured
as follows. In Sec.~II we explicitly show the LFWFs according to two holographic models.
In Sec.~III we concentrate on the pion wave function, discussing the adjustment of the
model parameters and distribution amplitudes, and in Sec.~IV we present some conclusions.

\section{Meson wave function in holographic models}

The comparison of form factors calculated both in the light-front formalism and in AdS
offers the possibility to relate AdS modes to LFWFs.
The main idea is that, with proper interpretation of certain quantities, the form factors
in both approaches can look similar and a mapping between the LFWF and AdS modes is
possible. For details see~\cite{BdT3,BdT4, VegaSchmidt3}. Here we only consider
two kinds of LFWFs obtained in two different holographic models (denoted by indices 1 and 2):

\begin{figure*}
  \begin{tabular}{cc cc}
    \includegraphics[width=2.0in]{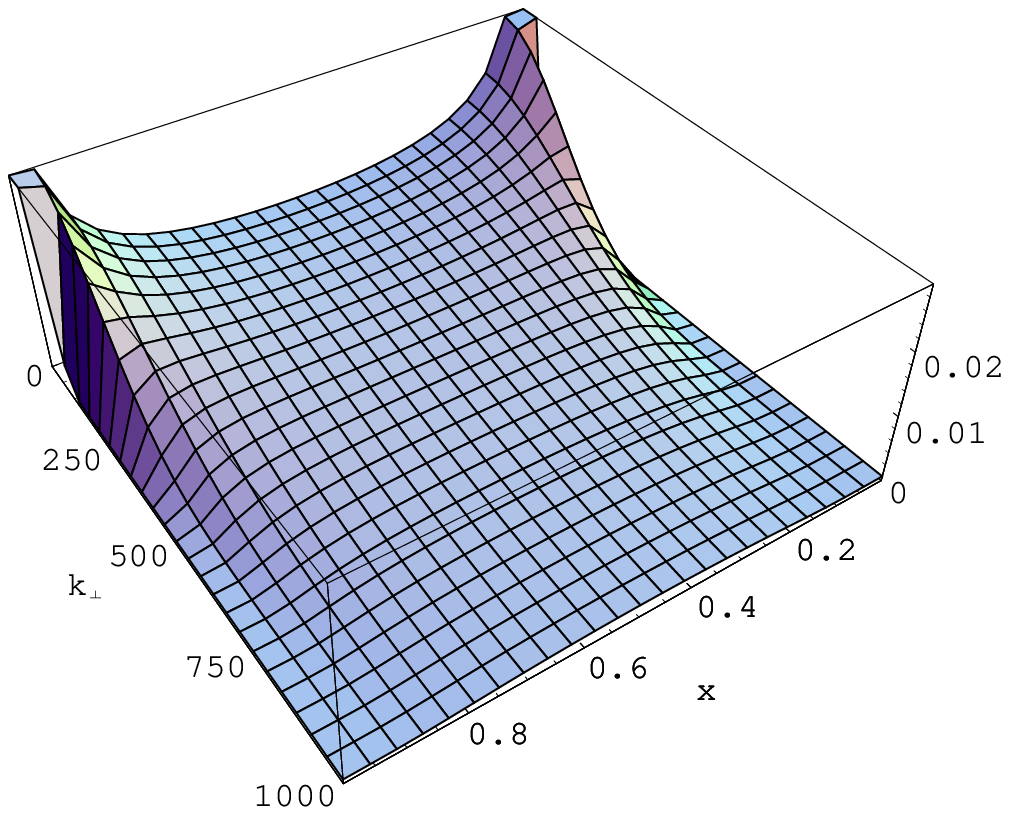} &
    \includegraphics[width=2.0in]{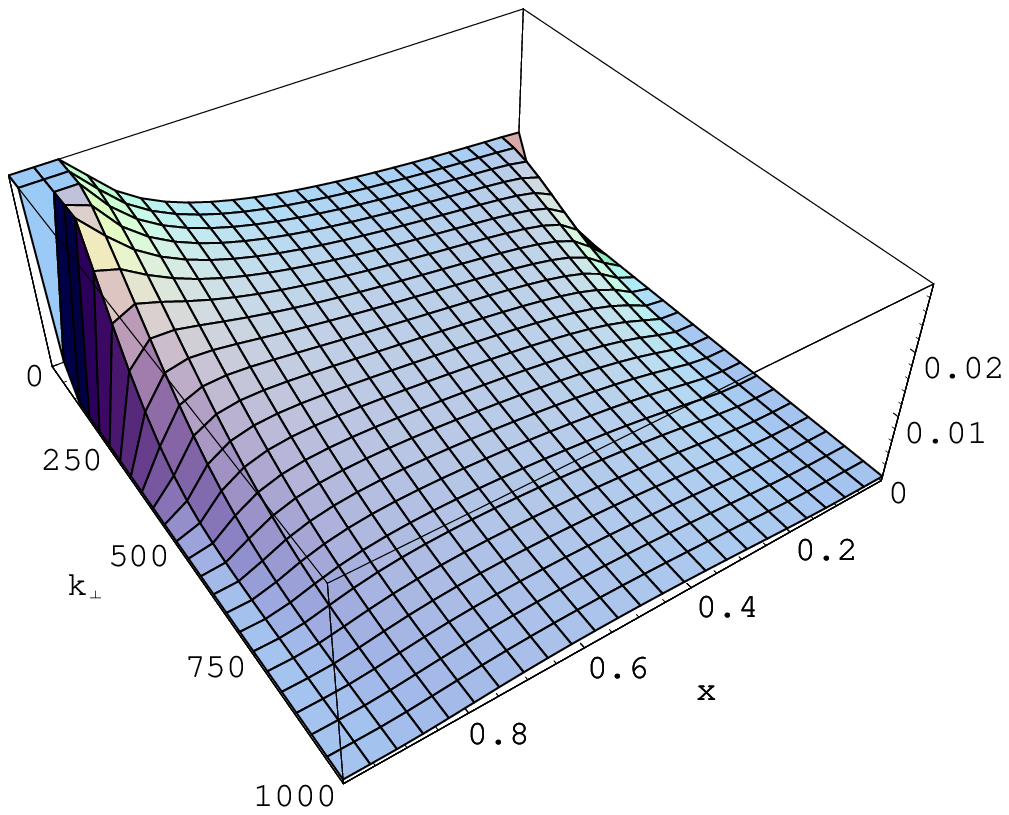}
  \end{tabular}
\caption{The pion wave function $\psi(x,k_{\bot})$ for $m = 4$ MeV. The
left graph corresponds to Eq.~(\ref{FnOnda1}) and the right one to
Eq.~(\ref{FnOnda2}). }
\end{figure*}

\begin{equation}
 \label{FnOnda1}
\hspace*{-.2cm}
 \psi_{q_1\bar q_2}^{(1)}(x,k_{\bot})
= \frac{4\pi A_1}{\kappa_1 \sqrt{x(1-x)}}
  \exp\biggl[-\frac{k_{\bot}^2}{2 \kappa_1^2 x(1-x)}
- \frac{1}{2\kappa^2_1} \biggl( \frac{m^{2}_{1}}{x} + \frac{m^{2}_{2}}{1 - x} \biggr)
 \biggr] ,
\end{equation}

\begin{equation}\label{FnOnda2}
\hspace*{-.2cm}
 \psi_{q_1\bar q_2}^{(2)}(x,k_{\bot})
= \frac{4\pi A_2}{\kappa_2} \sqrt{\frac{2}{1-x}}
 \exp\biggl[-\frac{k_{\bot}^2}{2 \kappa_2^2 x(1-x)}
-\frac{1}{2\kappa^2_2} \biggl( \frac{m^{2}_{1}}{x} + \frac{m^{2}_{2}}{1 - x} \biggr)
 \biggr] .
\end{equation}
In both cases massive quarks were included following a prescription
suggested by Brodsky and de T\'eramond~\cite{BdT5}.

\section{Example: The Pion}

The wave functions we consider depend on the parameters ($A_i$,$m_{1,2}$,$\kappa_i$) which
should be fixed.  We work in the isospin limit assuming that the masses of the $u$ and
$d$ quarks are equal: $m_u = m_u = m$. In this case we have a set of three
free parameters ($A_i$,$m$,$\kappa_i$) which is the same number of parameters
considered in other models~\cite{HMS}.

Using the quarks mass as an input, only two additional conditions are necessary.
These are related to the decay amplitudes for $\pi \rightarrow \mu \nu$ and
$\pi^{0} \rightarrow \gamma \gamma$~\cite{BHL}
\begin{equation}
 \label{Condition1}
 \int\limits_0^1 dx \int \frac{d^2 b}{16 \pi^{3}}
\psi_{q \overline{q}} (x,k_{\bot}) = \frac{F_\pi}{2 \sqrt{3}} \ \ \ \ 
{\rm and} \ \ \ \   
\int\limits_0^1 dx \, \psi_{q \overline{q}} (x,k_{\bot} = 0)
= \frac{\sqrt{3}}{F_\pi},
\end{equation}
where $F_{\pi} = f_\pi/\sqrt{2} \simeq 92.4$ MeV is the pion
leptonic decay constant.

\begin{figure*}
    \begin{tabular}{cc cc}
    \includegraphics[width=2.0in]{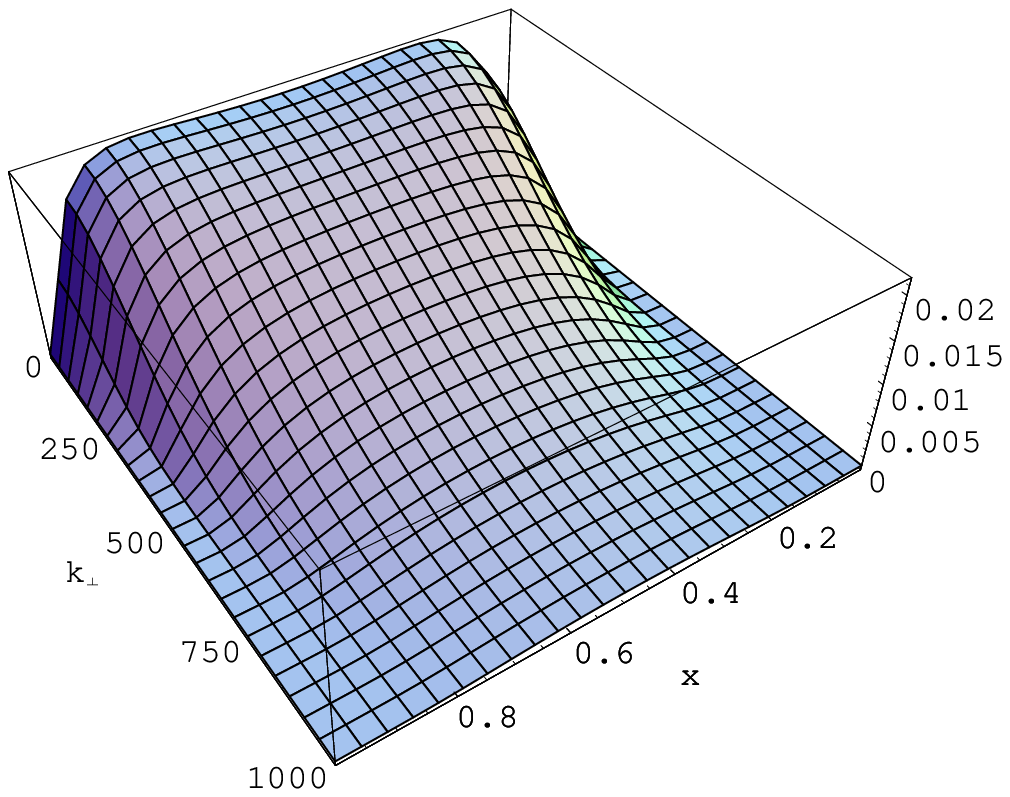}
  & \includegraphics[width=2.0in]{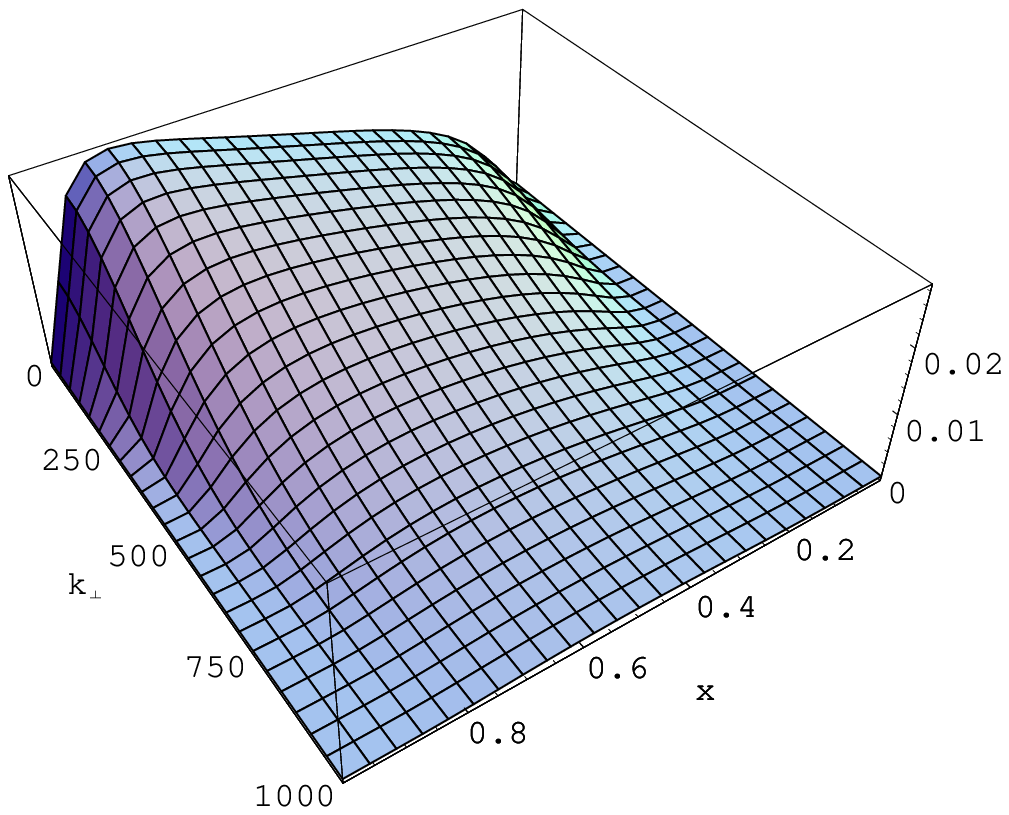}
    \end{tabular}
\caption{The pion wave function $\psi_{\pi}(x,k_{\bot})$ for $m = 330$
MeV. The left graph corresponds to Eq.~(\ref{FnOnda1}) and
the right one to Eq.~(\ref{FnOnda2}). }
\end{figure*}
Sometimes the average transverse momentum squared of a
quark in the pion, denoted as $\langle  k_{\bot}^2 \rangle_{\pi}$, with a value
of about 300 MeV$^2$~\cite{Metcalf} is used as an additional condition. Here we
use this further constraint to check the validity of the model.
\begin{table}
\caption{Parameters defining the LFWFs given by Eqs.~(\ref{FnOnda1}) and
(\ref{FnOnda2}) and predictions for $\sqrt{\langle k_{\bot}^2 \rangle_{q
\overline{q}}}$ and $P_{q \overline{q}}$.}
\begin{tabular}{| c | c | c | c | c | c | c |}
  \hline
Model & $\psi (x,k_{\bot})$ & $m$ (MeV) & $A$ & $\kappa$ (MeV) &
$\sqrt{\langle k_{\bot}^2 \rangle_{q \overline{q}}}$ (MeV) &
$P_{q \overline{q}}$  \\
  \hline
1  & $\psi_{1c} $ & 4   & 0.452  & 951.043 & 388.319 & 0.204 \\
  & $\psi_{1cs}$ & 330 & 0.924   & 787.43 & 356.478 & 0.279 \\
\hline
2  & $\psi_{2c} $ & 4   & 0.486   & 921.407 & 376.222 & 0.236 \\
  & $\psi_{2cs}$ & 330 & 0.965 & 781.218 & 353.877 & 0.299 \\
  \hline
\end{tabular}
\end{table}
The parameters $\kappa_{1,2}$, which are related to the Regge slopes,
enter in the holographic model
considered in Refs.~\cite{BdT4, VegaSchmidt3}.
Both quantities could in principle be fixed by
spectral data. Unfortunately, the pion mass is an exception since it
falls outside the Regge trajectories. Therefore, the values $\kappa_{1,2}$ have
been fixed by using the previous conditions.
The resulting pion wave functions $\psi_{\pi}(x,k_{\bot})$
for different values of the quark mass ($m=4$ and $330$ MeV) are displayed in Figs.1
and 2.

As an example we consider the meson distribution amplitude which is calculated
using~\cite{Lepage1}
\begin{equation}\label{phi_xq}
 \phi (x,q) = \int^{q^{2}} \frac{d^{2} k_{\bot}}{16 \pi^{3}}
\psi_{\rm val} (x, k_{\bot}).
\end{equation}
Fig. 3 shows the distribution amplitudes obtained using (\ref{FnOnda1}) and (\ref{FnOnda2})
in the previous expression. They are also compared to the prediction of PQCD with $\phi
(x, Q \rightarrow \infty) = \sqrt{3} F_{\pi} x (1-x)$ both for current and constituent
quark masses. As can be seen, an increase of the quark mass reduces the differences between
the two variants of the LFWFs.

\begin{figure}
  \begin{tabular}{cc}
    \includegraphics[width=2.0in]{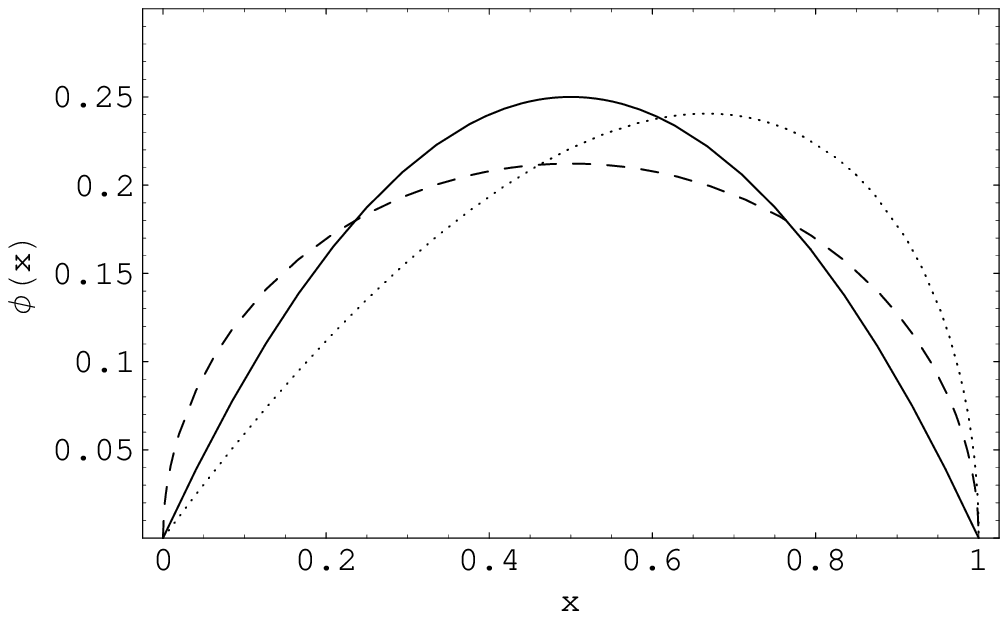}
  \end{tabular}
  \begin{tabular}{cc}
    \includegraphics[width=2.0in]{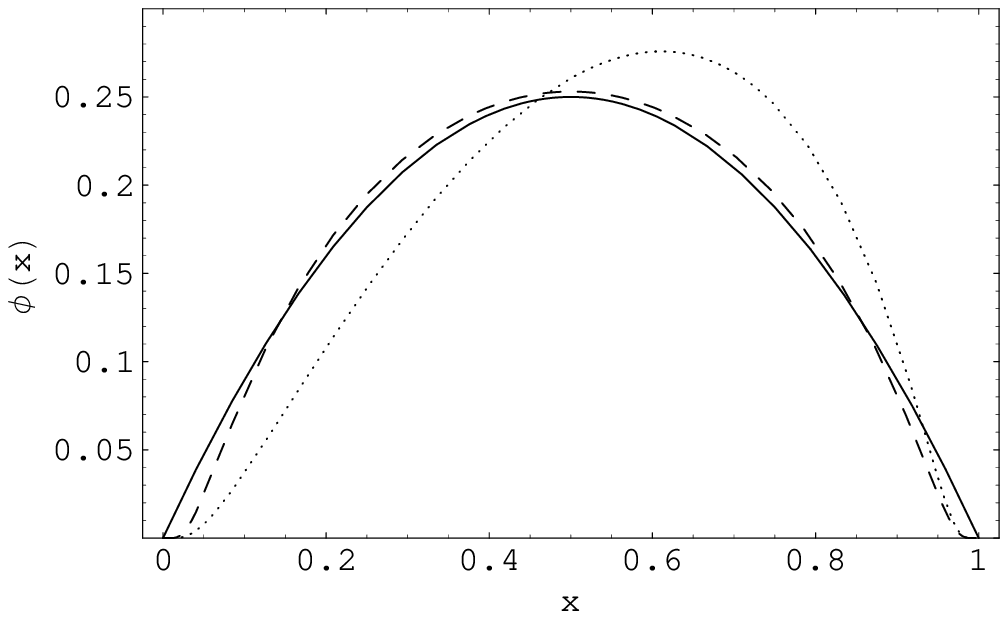}
  \end{tabular}
\caption{Pion distribution amplitudes using holographic LFWFs. Solid
lines correspond to the PQCD prediction, dashed lines
to model 1 and dotted ones to model 2, both for
$m = 4$ MeV (left panel) and for $m = 330$ MeV (right panel).}
\end{figure}

\section{Conclusions}

We have considered two kinds of wave functions for mesons in the
light-front formalism obtained by the AdS/CFT correspondence within
two soft wall holographic models. These wave functions have different $x$
dependence, which is less pronounced when the quark masses are
increased.
The parameters $\kappa_{1,2}$ used
in the holographic models can be fixed by spectroscopic data.
Taking quark masses as an initial input only one parameter remains (the
normalization constant $A_{1,2}$), which can be further fixed by the
normalization condition. Since pions do not really follow a Regge trajectory,
$\kappa_{1,2}$ must be fixed in a different way.
Due to the importance of the hadronic wave function in QCD the versions
considered here represent a clear example of the usefulness
of the AdS/CFT ideas in QCD.

\begin{theacknowledgments}
A. V. acknowledges the financial support from Fondecyt grants 3100028 (Chile). 
T.~B., T.~G. and V.~E.~L. 
were supported by the DFG under Contract No. FA67/31-2 
and No. GRK683 and the European Community-Research Infrastructure Integrating 
Activity ``Study of Strongly Interacting Matter'' (HadronPhysics2,
Grant No. 227431). 
\end{theacknowledgments}

\bibliographystyle{aipproc}   

\end{document}